# Simulation of DNA damage using Geant4-DNA: an overview of the "molecularDNA" example application

Konstantinos P. Chatzipapas[1], Ngoc Hoang Tran[1], Milos Dordevic[2], Sara Zivkovic[2], Sara Zein[1], Wook-Geun Shin[3], Dousatsu Sakata[4], Nathanael Lampe[5], Jeremy M.C. Brown[6], Aleksandra Ristic-Fira[2], Ivan Petrovic[2], Ioanna Kyriakou[7], Dimitris Emfietzoglou[7], Susanna Guatelli[8], Sébastien Incerti[1]

[1] Univ. Bordeaux, CNRS, LP2I Bordeaux, UMR 5797, F-33170 Gradignan, France

[2] Vinca Institute of Nuclear Sciences, National Institute of the Republic of Serbia, University of Belgrade, Mike Petrovica Alasa 12-14, 11351, Vinca, Belgrade, Serbia

[3] Physics Division, Department of Radiation Oncology, Massachusetts General Hospital & Harvard Medical School, Boston, 02114 MA, USA

[4] Division of Health Science, Osaka University, Osaka, Japan

[5] Unaffiliated, Melbourne, Australia

[6] Department of Physics and Astronomy, Swinburne University of Technology, Hawthorn 3122, Australia

[7] Medical Physics Laboratory, Department of Medicine, University of Ioannina, 45110 Ioannina, Greece

[8] Centre for Medical Radiation Physics, University of Wollongong, Wollongong, NSW 2522, Australia

## Abstract

The scientific community shows a great interest in the study of DNA damage induction, DNA damage repair and the biological effects on cells and cellular systems after exposure to ionizing radiation. Several *in-silico* methods have been proposed so far to study these mechanisms using Monte Carlo simulations. This study outlines a Geant4-DNA example application, named "molecularDNA", publicly released in the 11.1 version of Geant4 (December 2022). It was developed for novice Geant4 users and requires only a basic understanding of scripting languages to get started. The example currently proposes two different DNA-scale geometries of biological targets, namely "cylinders", and the "human cell". This public version is based on a previous prototype and includes new features such as: the adoption of a new approach for the modeling of the chemical stage (IRT-sync), the use of the Standard DNA Damage (SDD) format to describe radiation-induced DNA damage and upgraded computational tools to estimate DNA damage response. Simulation data in terms of single strand break (SSB) and double strand break (DSB) yields were produced using each of these geometries. The results were compared to the literature, to validate the example, producing less than 5 % difference in all cases.

## 1. Introduction

It is extensively recognized in the scientific community that the quantification of DNA damage induction and complexity in the cell, produced by charged particles, including heavy ions, is paramount [1-8]. Ionizing radiation influences several aspects of everyday life due to natural radiation background (cosmic and terrestrial sources), medical intervention (imaging and therapy), and space exploration (astronauts' missions and space tourism) [9]. Thus, it is important to achieve a better understanding of the fundamental mechanisms that are initiated by the interaction of radiation with living organisms.

At the subcellular scale, Monte Carlo simulations have proven to be useful for the quantification and assessment of radiation-induced damage, particularly to the DNA molecule. Several tools are currently available, such as PARTRAC [10], KURBUC [11], and RITRACKS [12] Geant4-DNA [13-16], gMicroMC [17, 18], and TOPAS-nBIO [19, 20], MPEXS-DNA [21], IDDRRA [22] that either are based on Geant4-DNA or include some of its features. A review of such tools can be found in [1]. They can simulate the physical interaction pathways of ionizing particles with biological media, also known as the physical stage, as well as the physico-chemical and chemical stages, where free radicals are produced by the deposition of energy in matter and interact with macromolecules such as DNA. Monte Carlo simulations and the comparison of their results with experimental (or real-life) data, allow to have an insight into the impact of each stage during an irradiation procedure. These *in-silico* tools are continuously upgraded to include new parameters and refine their models, in particular based on newly available empirical data.

This study presents the "molecularDNA" example of Geant4-DNA, which has been included in the *extended examples* category of Geant4 public release 11.1 (December 2022) [23-25]. "molecularDNA" is a user-friendly example for novice level Geant4 users that can be used to simulate radiation-induced damage at the DNA scale. It does not require more than a basic level of knowledge on scripting languages. This example currently includes two different DNA-scale designs of biological targets. The first design is a simple geometry called "cylinders", and, as the name suggests, it contains small DNA fragments placed in cells, modelled for simplicity as cylinders; its main purpose is to provide a simple setup for Geant4 regression testing of the functionality of the "molecularDNA" and of the Geant4-DNA physics capability. This implementation may also be used for small-scale DNA geometries, such as plasmids. The second design is a simplified human cell that may be used to investigate the impact at nanoscale of radiation environments of interest (i.e. during exposure to radionuclides, or in the context of the search of traces of life in the Solar system).

This public version of "molecularDNA", based on a non-public prototype by Lampe et al. [26, 27], has been significantly upgraded from its previous development iterations [28-34]. In particular, it includes the new chemistry model (Synchronous Independent Reaction Time model, so-called "IRT-sync") [35], the "*G4EmDNAChemistry_option3*" Geant4-DNA chemistry list based on Plante and Devroye [36], as well as an upgrade of computational tools to estimate DNA damage response (DDR) such as repair kinetics and survival probability of cells. In addition, in this study, simulation results are benchmarked against results available in the literature.

More extended information on the features of "molecularDNA" and its use may be found in https://geant4-dna.github.io/molecular-docs or in the Geant4-DNA website (http://geant4-dna.org).

## 2. Materials & Methods

### 2.1. Physical stage

The step-by-step modelling of physical interactions of primary and secondary ionizing particles with biological matter is named as the physical stage of Geant4-DNA simulations. These physical interactions may induce direct DNA damage, by directly depositing energy in DNA in the form of excitations and ionisations of the water molecule as well as solvated electrons. In this study, the damage is calculated individually for each event, which includes the tracking of a primary particle and of all its secondaries. The DNA geometry was defined as being composed of DNA materials, for which the current public version of "molecularDNA" does not include cross sections, meaning that no physical interactions are simulated within the DNA volumes. Cross sections of DNA materials will be included in future releases.

For simulating physical interactions, the Geant4-DNA option4 physics list [13] was utilized, based on previous studies by the Geant4-DNA collaboration. Option4 is the recommended physics list for studying physics interactions at the DNA level. Option4 physics models provide a more consistent treatment of the energy loss function (ELF) for liquid water, compared to option2 that was used in earlier studies [37]. Electrons and protons were simulated as primary particles that deposit energy inside the DNA, and cells (corresponding to the "cylinders", and the "human cell" geometry, respectively).

### 2.2. Pre-chemical and chemical stage

During the pre-chemical stage, the excited $H_2O^*$ and ionized $H_2O^+$ water molecules, which were produced at the physical stage, are dissociated into radical species based on dissociation channels used in [32] within 1 picosecond after the irradiation. Once the pre-chemical stage is completed, chemical species are diffused and react with each other or with the DNA targets to produce indirect damage [38]. The reaction rates and diffusion coefficients are incorporated in the default chemistry list of "molecularDNA", *G4EmDNAChemistry_option3*.

Currently, Geant4-DNA can simulate radiolysis in water that may include other substances like scavengers (e.g. oxygen concentration). However, in this approach, only the very early DNA damage is simulated, during which nucleotides that may be found very close to the track structure can be damaged by radical species. The simulated chemistry duration is short (less than 5 ns). During this period, most of the reactions happen between radiation-induced species, due to the high concentration of radiation spurs.

The chemical stage simulated in this work is based on the "IRT-sync" model that was recently released in Geant4 11.0 [35]. This model is an upgraded version of the prototype approach that combined the Independent Reaction Time (IRT) model with a maximum time step limitation and its implementation in Geant4 is described in [27-30, 33, 39, 40]. IRT-sync calculates a time step to the next chemical reaction using the IRT [35] method. At the end of the reaction, both the reactive products and the remaining molecules are considered to be synchronized when they are diffused for the next time step. By considering the diffusion of all species at the same time (synchronization process), IRT-sync provides the spatiotemporal information of the reactive species after each time step, which is used to calculate the interactions with the DNA geometry. IRT-sync is more efficient computationally, when compared to the step-by-step chemistry model, described in detail by Karamitros *et al.* [41, 42]. Because of its detailed analysis of chemistry interactions of all chemical species, the step-by-step model is extremely demanding in terms of computing time compared to the IRT and IRT-sync models.

The *G4EmDNAChemistry_option3* is a chemistry constructor implemented explicitly for IRT chemistry models. This model includes 15 molecular species, seven more when compared to the *G4EmDNAChemistry* constructor used for step-by-step radiolysis simulations. *G4EmDNAChemistry_option3* includes 72 chemical reactions, 63 more than what is available in *G4EmDNAChemistry*. These differences have been described in detail elsewhere [40].

### 2.3. Simulation Configuration

To benchmark the "molecularDNA" example with respect to previous studies of the Geant4-DNA collaboration, as well as to experimental data, this study kept the physical and chemical parameters as well as the damage definitions as close as possible to their previously established values. Histones were also considered for ensuring the completeness of the DNA nucleus geometry, as previously described in [28]. Histones tend to act as scavenging material, which means that they tend to absorb the reactive radicals. Therefore, it was important to take into consideration their impact.

### 2.3.1. Cylinders geometry

The "cylinders" geometry was inspired by Charlton et al. [43] and was first implemented by Nikjoo et al. [44]. Such simple geometry was implemented to explore how parameters that influence physical damage are related. It contains cylinders filled with DNA at random positions and directions. Using this design, users can calculate the number of radicals that are produced due to the interaction of radiation with water (i.e. radiolysis process). The "cylinders" geometry consists of a 3 µm radius water sphere filled with 200,000 individual 216 base pairs (bp) long straight DNA segments, as seen in Figure 1. Each segment is placed inside a cylindrical volume of 15 nm radius and 100 nm height [26].

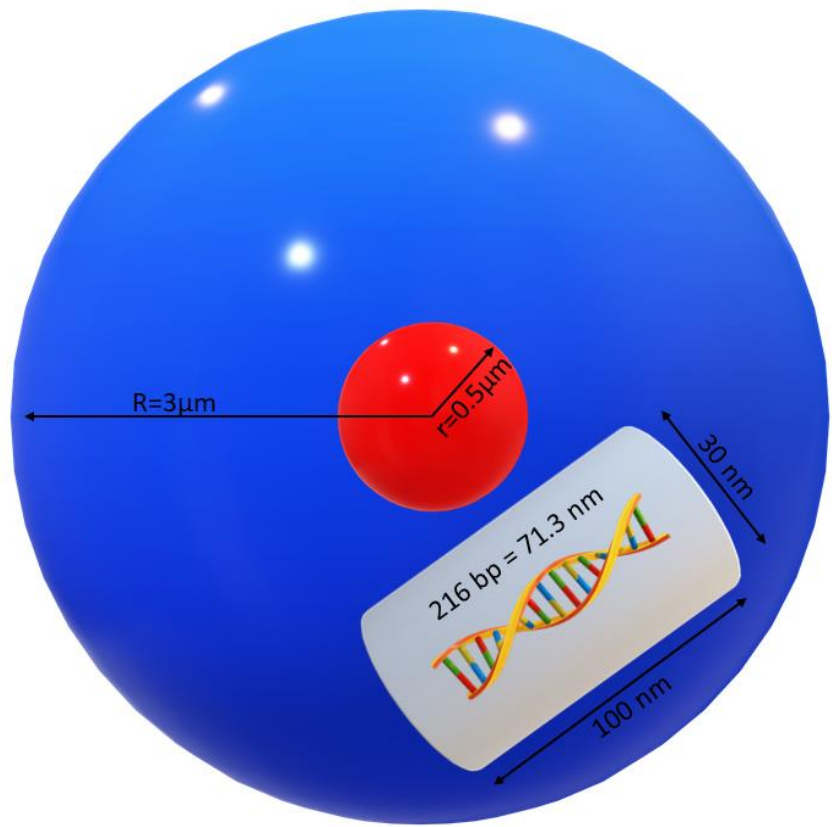

*Figure 1: Schematic representation of the "cylinders" geometry. A large water sphere (blue sphere in the figure) contains 200,000 cylinders including 216 bp of DNA as shown (in magnification) in the right-hand side corner. 4.5 keV electrons are randomly generated inside a sphere of 500 nm radius, as done in [26] (red sphere).*

Diffusion time was investigated for the durations of 1 ns and 1 µs. The maximum time step was set to 0.5 ns. All produced radicals were considered, without using any kind of particle production cut. Radical kill distance ($d_{kill}^{chem}$), which is the distance away from the DNA that the radicals are not tracked anymore to mimic scavenging, was set in the range of 0 to 9 nm. The range for direct interaction ($R_{dir}$), which is the maximum distance away from the DNA that energy deposits count as DNA damage, was set to 7 Å. To produce direct damage, the minimum ($E_{min}^{break}$) and maximum ($E_{max}^{break}$) energy thresholds, which refer to the accumulated energy in a specific area (e.g. a sugar, or a sphere/cylinder) of the DNA, were both set to 17.5 eV. Electrons of 4.5 keV kinetic energy were utilized to irradiate the spherical world containing the cylinders. They were randomly generated inside a sphere of 500 nm radius, as done in [26] (red sphere in Figure 1). The Geant4-DNA option4 was used as physics list. These parameters are consistent with those from previously published data and can be found in Table 1 [26].

*2.3.2. Human cell geometry*

The final set of simulations focused on using a geometrical model of a human cell ("human cell"). This part of the study focused on regression testing with a previous version of the example, described in detail in [30]. More specifically, the implemented geometry consists of a continuous Hilbert curve fractal-based DNA chain composed of straight and turned chromatin sections, including histones [27]. The continuous DNA chain structure is approximately 6.4 Gbp long and is placed inside an ellipsoid of 7.1 μm x 2.5 μm x 7.1 μm semi-axes that imitates the human cell nucleus (Figure 2). This setup results in an effective nucleus density of ~0.015 bp/nm$^3$ [45]. The Geant4-DNA option4 physics list was selected for calculating electron and proton energy transport, using a 40 % probability cut for the DNA strand break induction by OH$^\bullet$. The 40 % probability cut to induce a strand break was chosen based on the assumption made in earlier studies, which considered that ~13 % of all reactions between DNA and OH, induce a double strand break, which has been proposed in [46, 47]. A maximum diffusion time of 5 ns (equivalent to 9 nm diffusion distance) was implemented. Histones were considered as perfect scavengers, which translates to killing free radical species that enter the histone region.

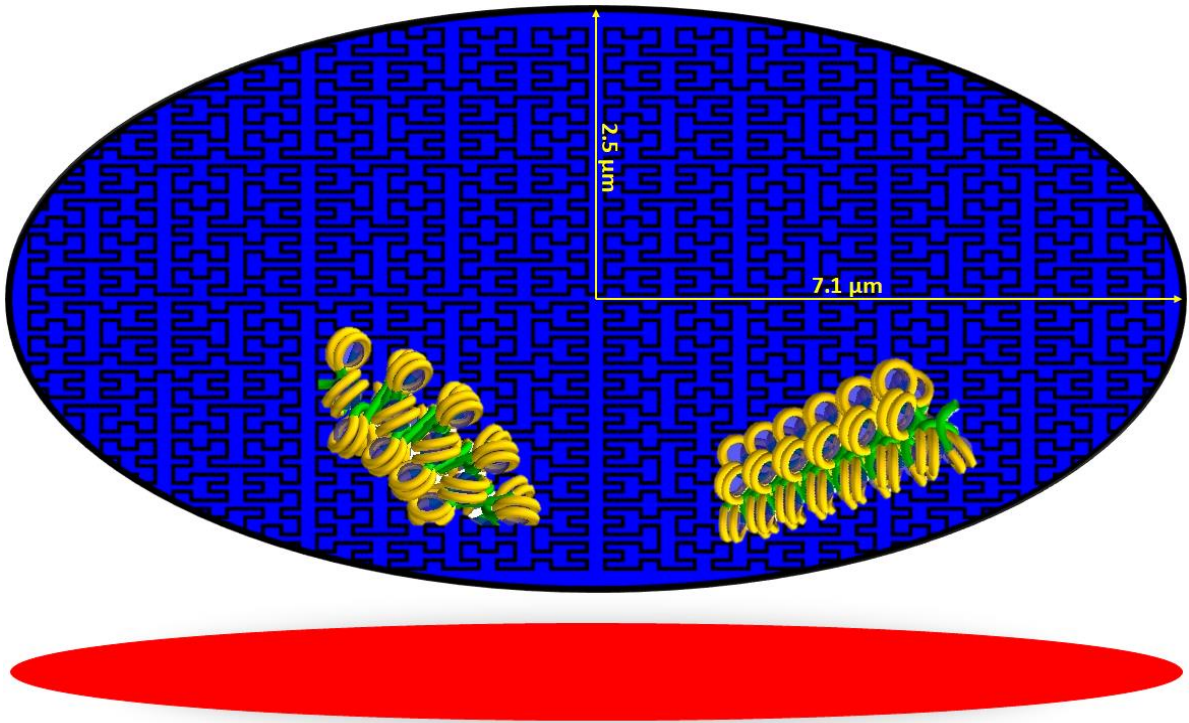

*Figure 2: Simplistic 2D representation of the "human cell" geometry. The black lines inside the ellipsoid water cell represent the fractal path of DNA (Hilbert curve). Two segments of chromatin fibers including histones are shown, a straight one as well as a turned one (not in scale with the rest of the geometry in the figure). The red disk represents the source of radiation.*

The results of this simulation study were compared with both experimental [48-57] and other simulation data [28, 30, 31, 33], in terms of calculation of the number of strand breaks (SB) and of direct/indirect damage. To generate a direct DNA damage lesion, the energy deposited during a physical interaction should be closer than 3.5 Å to a DNA base ($R_{dir}$). This distance corresponds to the radius of nucleotide molecules and takes into account the charge transfer effect in the hydration shell [30]. The generation of a direct DNA damage lesion is also energy dependent, and a linear probability model was implemented over an energy range ($E_{min}^{break}$ to $E_{max}^{break}$) of 5 eV to 37.5 eV (0 to 100% change), in line with the work of Lampe et al. [27]. The parameters utilized in the "human cell" simulation are summarized in Table 1.

Table 1: Parameters used in this study during the simulations of the "human cell" geometry, compared to previous studies, as well as to the "cylinders" geometry.

|  | KURBUC | PARTRAC | GEANT4-DNA_2020/ THIS STUDY | CYLINDERS |
|---|---|---|---|---|
| $R_{dir}$ (Å) | 1.7-3.25 | 2VDWR | 3.5 | 7 |
| $E_{min}^{break}$ (eV) | 17.5 | 5 | 5 | 17.5 |
| $E_{max}^{break}$ (eV) | 17.5 | 37.5 | 37.5 | 17.5 |
| $p_{OH}^{break}$ | 0.13 | 0.7 | 0.405 | 1 |
| $T_{chem}$ (ns) | 1 | 10 | 5 | 1-1000 |
| $d_{kill}^{chem}$ (nm) | 4 | 12.5 | 9 | 0-9 |

VDWR stands for the atomic van der Waals radius (1.2, 1.7, 1.5, 1.4, 1.9 Å for H, C, N, O, P respectively). During PARTRAC simulations, to adjust for cross-section of molecules and consider hydration shell, VDWR was multiplied by 2.

### 2.4. Damage classification

The simulation quantifies several types of DNA damage according to the level of complexity and the interaction type (i.e. physical or chemical). The damage classification scheme of Nikjoo et al. [44] is adopted; single strand breaks (SSB), double strand breaks (DSB), their damage complexity (denoted as $SSB_+$, $DSB_+$, and $DSB_{++}$), and the source of damage (namely direct, indirect, mixed, and hybrid) are calculated. This information may be used to investigate different processes/yields of DNA damage induction and enables the user to focus on specific damage repair mechanisms of interest.

A DSB is defined as two SSBs on two opposite DNA strands within a distance of 10 bp, with two additional complexity classifications, namely DSB-plus ($DSB_+$) and DSB-plus-plus ($DSB_{++}$). A $DSB_+$ requires a DSB and at least one additional break within 10 bp, while a $DSB_{++}$ requires at least two DSBs along the chromatin fiber segment. During post-processing analysis, the DNA segment length may be optimized to produce results comparable to other studies; in Geant4-DNA, we consider a default length of 100 bp.

Regarding the complexity classification, which is based on the source of the damage (direct or indirect), a DSB is classified as DSB hybrid ($DSB_{hyb}$) if a DSB contains both direct and indirect damage. If both direct and indirect breaks are included in a complex DSB ($DSB_+$ or $DSB_{++}$), it is classified as DSB mixed ($DSB_{mix}$). All other simple DSBs are classified as DSB direct ($DSB_d$) or DSB indirect ($DSB_i$).

To be more accessible to the community, "molecularDNA" can now produce DNA damage simulation results in the recently proposed Standard DNA Damage data format (SDD) (see Schuemann et al. [58]). In this sharable datafile format, "molecularDNA" accumulates information regarding the radiation source, the target object, and the simulation environment, based on the guidelines set in [58]. Additionally, the SDD file includes information about the damage type (single or double strand break), its spatial position, and its complexity, as produced after processing every single simulation event. A recent update to the SDD file has been included in the "DSB type" field to store information important for repair and survival estimation models. This update extends the scale of damage type, which in the original version is either 0 or 1, meaning whether a DSB exists or not. "molecularDNA" further extends this scale up to 5, where 1 stands for DSB-direct damage, 2 stands for DSB-indirect damage, 3 stands for DSB-hybrid damage (i.e. including both direct and indirect breaks), 4 stands for $DSB_+$ and 5 stands for DSB++. Users may use either the output from the simulation in ROOT file format [59], or the SDD file format, or even both of them, to save their results.

## 2.5. Repair model

The Geant4 11.1 (Dec. 2022) version of the "molecularDNA" example will include a DNA damage repair model for irradiated cells, based on the mathematical framework that was proposed by Belov et al. [60], and was further investigated with Geant4-DNA by Sakata et al. [30]. This model calculates the accumulated repair protein yield through the consideration of four principal DSB repair pathways. More analytical discussion on repair pathways can be found in the literature [61]. These pathways are:

  a. The non-homologous end-joining (NHEJ),
  b. The homologous recombination (HR),
  c. The single-strand annealing (SSA), and
  d. The alternative end-joining mechanism (Alt-NHEJ).

This model takes into account both the number of repairable ($N_{ncDSB}$) and non-repairable ($N_{cDSB}$) DSBs per unit of dose and per cell (DSB/Gy/cell). The rate of the population change of the repair proteins can be mathematically expressed by equation (1):

$$\frac{dN_0}{dt} = a(L)\frac{dD}{dt}N_{cDSB} - V_{NHEJ} - V_{HR} - V_{SSA} - V_{microSSA} \qquad (1)$$

$N_0$ is the total number of DSB ($N_{ncDSB} + N_{cDSB}$), which is proportional to the population of repair proteins. The parameters $V_{NHEJ}, V_{HR}, V_{SSA}, and\ V_{microSSA}$ characterize the way that each repair model affects the remaining repair proteins. D is the dose (Gy). $a(L)$ describes the DSB induction per unit of dose in each cell (Gy$^{-1}$ cell$^{-1}$) and depends on the Linear Energy Transfer (LET).

The repair model described in equation (1) utilizes the mass-action kinetics approach to simulate the DNA Damage Response (DDR) by repair proteins. This equation describes the balance between the increasing damage term ($a(L)$), compared to the decreasing following terms ($V_{NHEJ}, V_{HR}, V_{SSA}, and\ V_{microSSA}$). This model is based on two major assumptions:

1. The sample is irradiated at a clinical dose rate (Gy/s), and
2. At t=0, which corresponds to the moment right after the irradiation, the dose has been already deposited to the target, meaning dD/dt=D with no new DNA damage induced after this initial time point.

It is important to state that most of the rate constants are evaluated by fitting to the experimental data on the kinetics of different stages of DSB repair [60]. By solving these differential equations, the model may estimate the yield of the target repair proteins as the system evolves in time (this study focused on the yield of γ-H2AX foci). γ-H2AX is involved in the steps leading to chromatin response after DSBs have been produced. Because of that, γ-H2AX may be used as a biomarker of radiation exposure, and, if correlated with time, the number of remaining foci reproduces the repair time curve. More information about this may be found in [30, 62]. In this study DSB+ and DSB++ are considered to be non-repairable damage, and they are calculated using the following assumption: $N_{cDSB} = N_{DSB+} + 2 \cdot N_{DSB++}$.

Python was used to apply this model to the results of the "molecularDNA" in post-processing, after the simulation is completed. This tool, without any modification, can also be applied by users to data found in the literature, independently from the simulation environment. Python was used for two main reasons. It is a user-friendly Object-Oriented language that can be easily used, and, most importantly, can be modified to meet the needs of the user. Additionally, there is no need for installing other software or libraries to solve differential equations. This translates to the fact that the tool can easily be used in any computer system.

### 2.6. Cell Survival Function

To estimate the clonogenic surviving fraction (SF) of specific cells, the two-lesion kinetics (TLK) model proposed by Stewart et al. [63] was utilized in this study. The TLK model includes kinetic processes of fast- and slow- DNA repair, and, based on lethal DNA damage, it can calculate the SF of a cell population. Both repair mechanisms consider simple rejoining of damaged base pairs at the same position and are described by the term L(t). This model also includes multiple-lesion repairs (second-order repairs), which are expressed by $L^2$(t). It must be stated that multiple-lesion repairs mechanism may lead to complex aberrations due to incorrect rejoining.

The TLK model consists of a set of equations, which are presented in equations (2), (3), (4).

$$\frac{dL_1(t)}{dt} = D(t)Y\Sigma_1 - \lambda_1 L_1(t) - \eta L_1[L_1(t) + L_2(t)] \quad (2)$$

$$\frac{dL_2(t)}{dt} = D(t)Y\Sigma_2 - \lambda_2 L_2(t) - \eta L_2[L_1(t) + L_2(t)] \quad (3)$$

$$\frac{dL_f(t)}{dt} = \beta_1 \lambda_1 L_1(t) + \beta_2 \lambda_2 L_2(t) + \gamma\eta[L_1(t) + L_2(t)]^2 \quad (4)$$

These equations include several parameters, namely:

a. Repair probability coefficients, which represent the rate of rejoined lesions (λ and η), and
b. Lethality probability coefficients, which represent the probability that a residual lesion may lead to cell death (β and γ).

To validate the newly developed algorithm, this study implemented the same parameter values adopted in [31].

Additionally, $L_1$(t) is the number of lesions per cell in the fast-repair process at a given time t after the beginning of the irradiation procedure. $L_2$(t) is the number of lesions per cell in the slow-repair process at a given time t. $L_f$(t) is the number of lethal lesions that may lead to cell death at time t. D(t) is the dose rate, Y is the size of the cell in Giga base pairs (Gbps). $\Sigma_1$ corresponds to the number of simple DSB, while $\Sigma_2$ corresponds to the number of irreparable damage (complex DSB - $N_{cDSB}$).

$\lambda_1$, $\lambda_2$, and η correspond to fast-, slow-, and binary-rejoining processes, respectively (expressed in $h^{-1}$). Similarly, $\beta_1$, $\beta_2$, and γ correspond accordingly to each rejoining process. As in the previous studies [31, 63], $\beta_1$ was set to 0, because simple DSBs do not affect cell survival more than 2 weeks after the irradiation session. After integrating the produced yields, the SF is calculated by:

$$SF(t) = ln(-L_f(t)) = ln\left(-\int_0^t (\beta_1\lambda_1 L_1(t) + \beta_2\lambda_2 L_2(t) + \gamma\eta[L_1(t) + L_2(t)]^2)dt\right) \quad (5)$$

This study assumes that the SF is calculated 336 h after irradiation, as followed in the experimental procedure [64], which considered that any colony with more than 50 cells, after a 14-day incubation in 5% $CO_2$ incubator at 37 °C, had survived. D(t) is set to 60 Gy/h until the target dose is delivered, and the time step of the integration is set to $1\cdot10^{-2}$ hours, following the methodology of Sakata et al. [31]. For this public release, the differential equation is solved numerically via the fourth-order Runge-Kutta method using the Scipy Python library [65]. These libraries are already included in the Python default library. This model, implemented in Python, can be used with any type of data, regardless of the way that they have been produced or their format, meaning that it may be used independently from the Geant4-DNA simulation environment.

## 3. Results and Discussion

Figure 3 shows the number of radicals, SSB and DSB, produced as a function of the radial kill distance, plotted for different end times (1 μs and 1 ns), in the "cylinders" geometry, when irradiated with 4.5 keV electrons. The results are obtained using the default chemistry list of the "molecularDNA" (IRT-sync) and one documented in Lampe et al. [26]. $10^5$ histories were simulated to obtain a statistical uncertainty lower than 0.01 %.

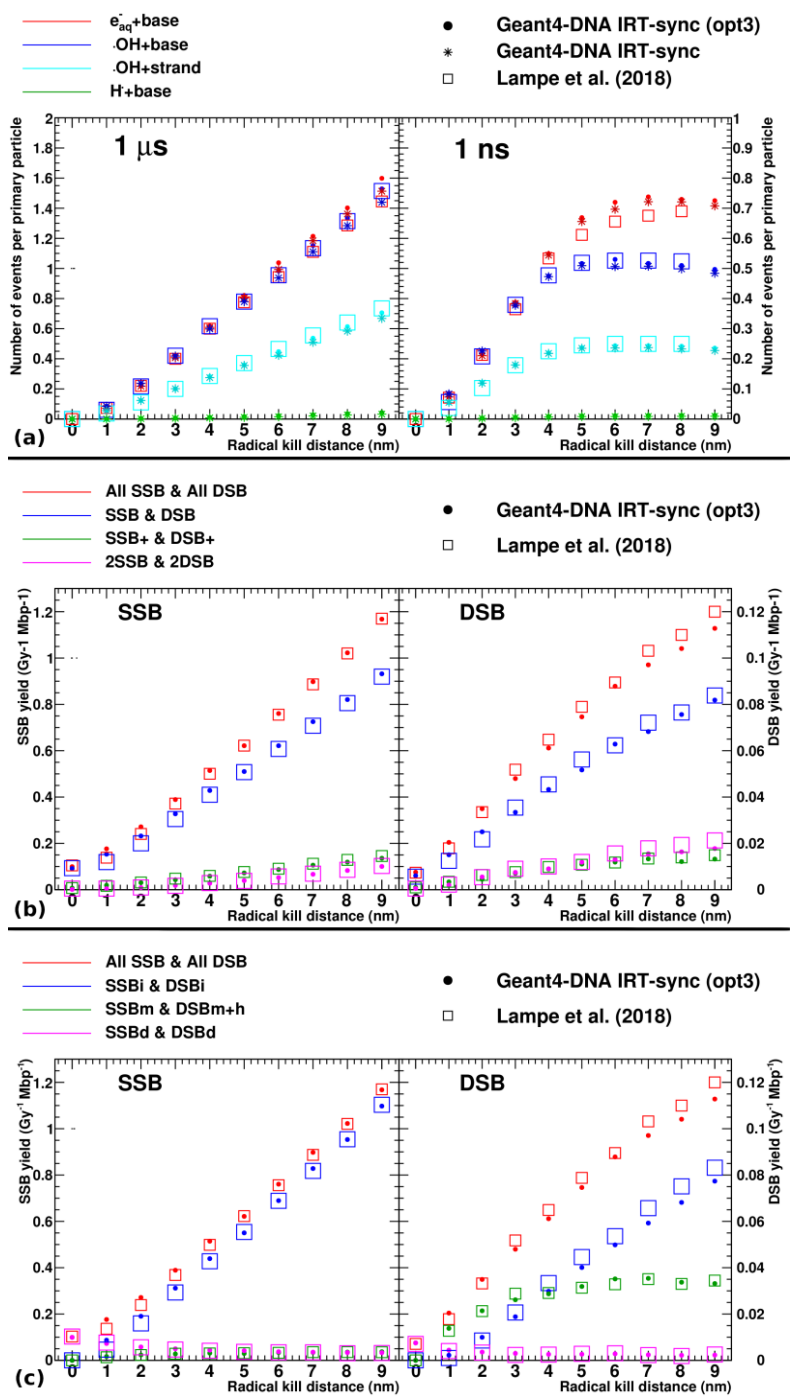

*Figure 3: a) The number of chemical reactions produced as a function of the radial kill distance for incident 4.5 keV electrons, when the simulation stops at 1 μs (left) and at 1 ns (right) using both the default chemistry list that is included in "molecularDNA" (opt3- G4EmDNAChemistry_option3) and the chemistry list used in Lampe et al. [26] (G4EmDNAChemistry). b) SSB and DSB yields obtained using the default chemistry, c) SSB and DSB yields plotted considering the source of the damage using the default chemistry, for "cylinders" geometry.*

The first set (a) presents the difference observed in the number of induced radicals between three different setups. The first setup (squares) is the data published by Lampe et al. [26] and act as reference. The second setup (stars) represents the data produced using "molecularDNA" together with the *G4EmDNAChemistry* constructor, which is the one used in [26], to have a direct comparison with the reference. The third setup (dots) represent the data produced by "molecularDNA" using the *G4EmDNAChemistry_option3*, which is the default chemistry list for "molecularDNA", released in the 11.1 version of Geant4. No significant difference between the three different cases is observed.

The last two plots (b and c) show the SSB and DSB yields, using the default chemistry list of "molecularDNA, while considering the complexity of the damage, as well as the source of the damage (direct and indirect), respectively. The maximum statistical difference between the results set obtained with the two chemistry physics list was found to be 5 %, which is deemed to be a satisfactory agreement.

The next step of this study was the investigation of the damage that may be induced to human cells by ionizing radiation, in the "human cell" geometrical configuration. The human cell geometry consists of an ellipsoid cell nucleus (semi-axes a=7.1 μm, b=2.5 μm, c=7.1 μm) and DNA of ~6.4 Gbp length [66]. The "human cell" was irradiated using protons with energies in the range of 0.15 to 66.5 MeV, corresponding to the range of LET from 73.5 to 1 keV/μm (i.e. 73.5, 62.5, 47, 36.7, 28, 26, 24, 19.5, 14.5, 9.6, 5, 2.5, 1 keV/μm). Both SSB and DSB yields, as well as their ratio were calculated. Figure 4 shows the DSB yield and its comparison to previously published data. Error bars have been included in the size of the bullet or line, as they were smaller than 0.01 %. The comparison of the SSB/DSB ratio calculated in the context of this study, against previously published data is presented in Figure 5. Simulation results produced the same curve trend as other studies where the DSB yield is in direct relation with LET of incident particles. As expected, when the LET of incident particles increases, particle interactions as well as the consequently produced free radicals are denser, resulting in denser particle/radical-DNA interactions, which produce a higher probability for DSB occurrence.

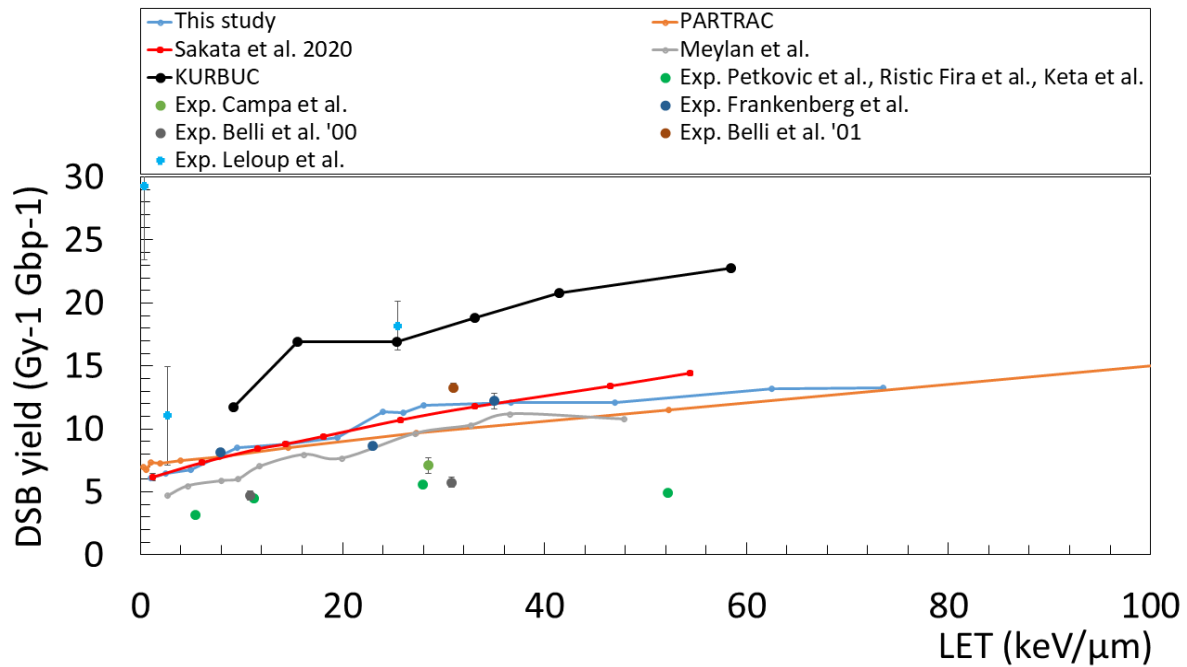

*Figure 4: DSB yield for incident protons as a function of LET, for "human cell" geometry. Both simulation and experimental data are included [30, 48-57].*

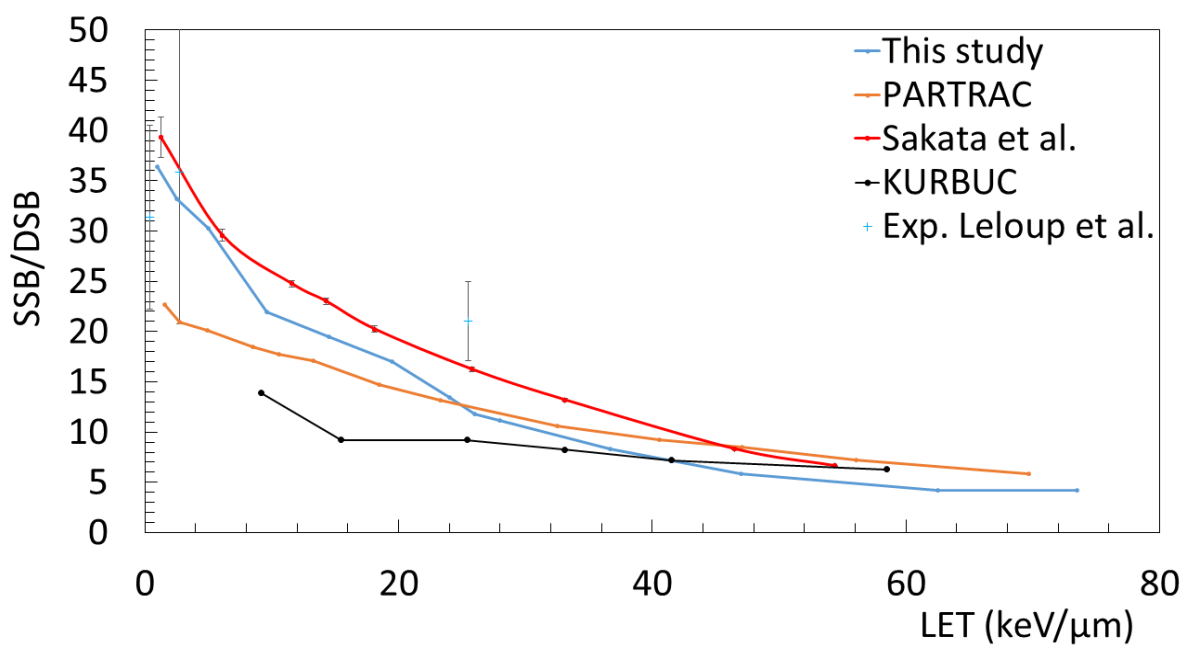

*Figure 5: Ratio SSB/DSB for incident protons as a function of LET, for "human cell" geometry. Both simulations and experimental data are included [30, 54].*

Furthermore, this is reflected in the SSB/DSB ratio of Figure 5 where SSBs have a higher probability of regrouping into DSB, lowering the ratio as the LET of incident particles increases. However, for very low LET, no significant influence was observed on the SSB yield as a function of LET. The results of this study are consistent with data previously published by the Geant4-DNA collaboration [28, 30, 33], as well as experimental data available in the literature.

"Human cell" was also implemented in a simulation that produced an accumulated absorbed dose of 100 Gy. To this purpose 1 MeV protons irradiated the target. Figure 6 shows the fragments' length distribution frequency obtained with the updated "molecularDNA", against available data including experimental data found in the literature, PARTRAC simulation data, as well as previously published data by the Geant4-DNA collaboration. Fragments are defined as broken segments of a DNA chain. Uncertainties are included in the size of the bullet or line. It should be noted that this study utilized the same example ("molecularDNA"), geometry, radiation field and set of parameters of Sakata et al. [30]. This regression test shows a mean difference in the results of 2% between the old and new version of the extended example (higher than the simulation statistical uncertainty of 0.1 %), due to the different chemistry model that was adopted (IRT-sync here and prototype IRT in Sakata *et al*).

The statistical uncertainty of DSB yield calculated by this study's simulations is less than 0.1 %. These differences are the result of the different chemistry model (IRT-sync), as well as the use of a different chemistry list during the simulations.

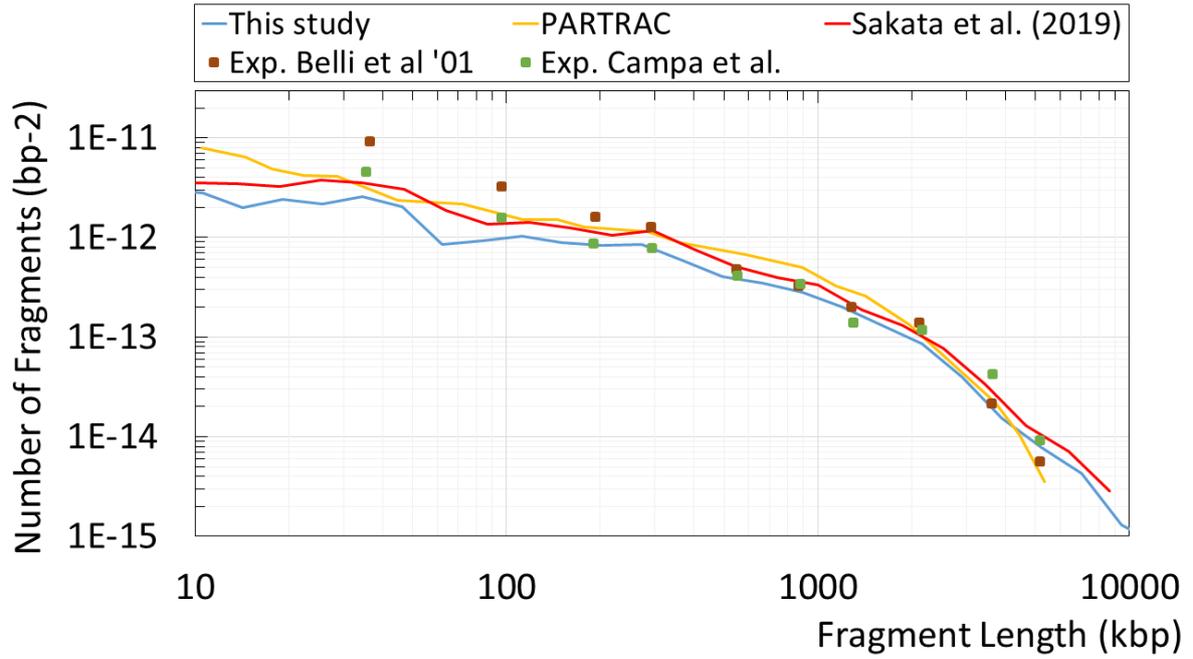

*Figure 6: Histograms of the fragment length frequency distribution for 1 MeV incident protons. The results of simulation (lines) are overlaid with the experimental measurements (squares) [30, 53, 55]. Statistical uncertainties are within the size of the bullet or line.*

In the context of this study, an additional *in-silico* tool (that is included in "molecularDNA") was developed in Python, which calculates DNA damage kinetics for the γ-H2AX type of cells using the NHEJ model, based on the equations proposed by Belov et al. [30, 60], shown in Figure 7 together with previously published data, based either on *in-vitro* or *in-silico* studies. The γ-H2AX yield was plotted as a function of time after irradiation, up to 25 h. In this specific case, the results were obtained with $10^5$ incident electrons of 0.662 MeV kinetic energy. Modified parameters used in our implementation are shown in table 2. The rest of the parameters have been set as presented in [60].

*Table 2: Modified values used in the implementation of the repair model in the context of this study. The complete list of the values can be found in [60].*

| PARAMETER | VALUE | |
|---|---|---|
| *a* | 30.5 | DSB/Gy/cell |
| $N_{ir}$ | 0.115 | |
| $K_1$ | 11.052 | $M^{-1} \cdot h^{-1}$ |
| $K_8$ | 0.1932 | $h^{-1}$ |
| $P_2$ | 0.39192 | $h^{-1}$ |

Results were compared to the experimental data obtained in the case of normal human skin fibroblasts (HSF42) exposed to gamma rays emitted by a $^{137}$Cs source, for a 1 Gy dose [52, 67]. They are in good agreement with data previously published by the Geant4-DNA collaboration and with experimental data. The calculated γ-H2AX yield with optimized rate constants leads to a very good agreement with the experimental data (within a 1.6 % difference on average). The decrease of the γ-H2AX yield matches reasonably well the experimental data.

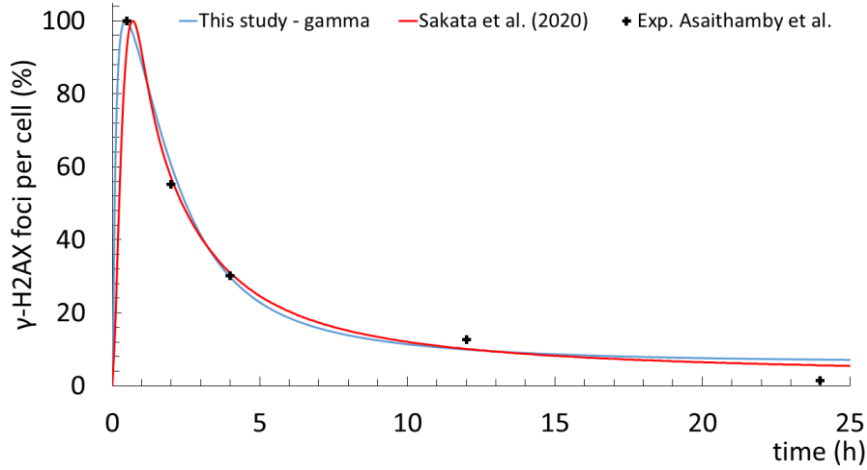

*Figure 7: Damage kinetics presented as γ-H2AX yield as a function of repair time, using Belov et al. differential equations for the NHEJ model [30]. Normal human skin fibroblast cells (HSF42) were irradiated with $^{137}$Cs, for an absorbed dose of 1 Gy, as described in [67].*

In Figure 8, the survival function of the cell is plotted against the absorbed dose, under irradiation with 70 keV protons. Data by the literature was used to validate our new implemented tool for the calculation of the survival of cells. More specifically the early DNA damage is summarized in table 3. It must be stated that the experimental and simulation setup was constituted by the parallel proton source, a PMMA block as well as the target being the cell DNA. A type of normal human skin fibroblast cells, named NB1RGB (No. RCB0222) and distributed by the RIKEN BioResource Center Cell Bank (RIKEN, Japan) was used as reference. A very good agreement is observed between the new "molecularDNA" simulation and the previously published results. Compared to data published by the Geant4-DNA collaboration in [31] no difference was observed, while with experimental data published in [64] the average difference is less than 6 %. To note, the same input parameters were adopted here and in [31] (see Section 2.6).

*Table 3: Damage data used with the survival model.*

|  | PMMA 0 MM | PMMA 32 MM |
|---|---|---|
| **SIMPLE DSB YIELD ($GY^{-1}$ $MBP^{-1}$)** | 4.11 ± 0.14 | 4.69 ± 0.17 |
| **COMPLEX DSB YIELD ($GY^{-1}$ $MBP^{-1}$)** | 0.74 ± 0.11 | 1.04 ± 0.12 |

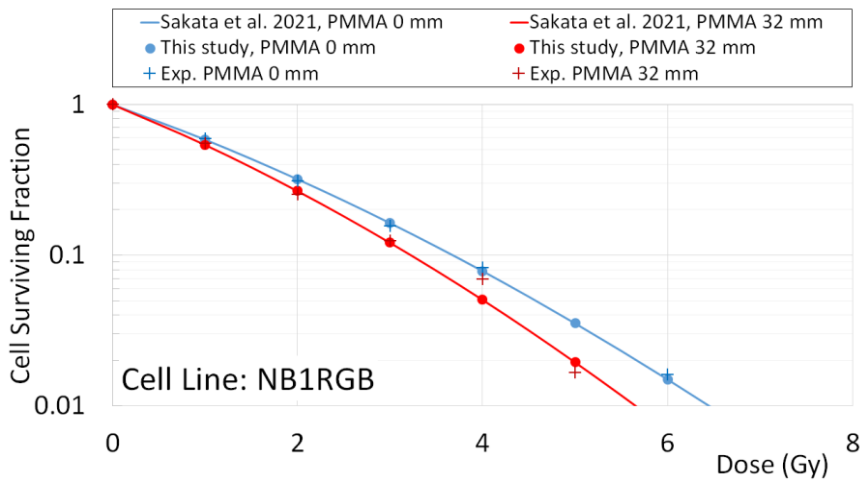

*Figure 8: Surviving fraction of NB1RGB cell line as a function of delivered dose [30].*

## 4. Conclusion

This is the first time that "molecularDNA" is made publicly available in Geant4, since the release 11.1 (December 2022). This simulation evolved in time since its first prototype [26, 27]. Novel features include the adoption of the IRT-sync chemistry model [35], the adoption of SDD data format, as well as tools to further investigate the DDR of cells. We have presented in detail the way that "molecularDNA" may be used by interested users to produce data on DNA damage induction, foci repair and survival fraction.

To summarize, this study evaluated the Geant4-DNA "molecularDNA" example against data available in the literature, including previous versions of the example [29-31] for regression testing purposes. "molecularDNA" is particularly suitable for novice users of Geant4 to establish complex multilevel studies, starting from the modeling of simple DNA molecules such as the DNA included in the "cylinders" geometry, up-to the level of complexity found in the "human cell" geometry; followed by the simulation of the irradiation of these geometries and eventually analyzing their outcome and the response of living cells to the produced damage. "molecularDNA" example may produce data on DNA damage quantification taking into consideration both physical and chemical interactions, namely direct or indirect damage. The implementation of IRT-sync made the "molecularDNA" significantly faster when compared to previous versions and other available examples that use the step-by-step chemistry model. The user may also adopt the SDD datafile format to estimate the complexity of the damage and/or to share the simulation output.

Furthermore, the simulation output may be analyzed not only to calculate the number of SSB, DSB, complex damage and DNA fragments, but also to study the way that a cell responds to early DNA damage response (DDR). Response may be calculated in terms of damage kinetics based on specific models (NHEJ, HR, SSA, Alt-NHEJ), as well as in the more commonly used quantity of survival function of cells. The survival function, in general, has been used in clinical practice to investigate the way that radiation therapy is applied, as it is important to estimate the probability of cells to survive after one or multiple irradiation procedures.

It is important to remind that the tools, which were developed to calculate damage kinetics as well as the survival function of cells, can be used independently from the Geant4 simulation. Thus, they can be applied universally to experimental studies or other *in-silico* methods. They can be applied directly to SDD files produced in any way, following the SDD guidelines [58]. SDD files may also be used to be applied on different models, such as the Medras, which has been described and implemented in [68, 69]. A future work will compare models described in this article with other models, including Medras and NanOx[70].

"molecularDNA" is an original complete "chain" from the fundamental physics interactions to the biological end-points, for specific cell lines, dose, and dose rates. Only a basic knowledge of any scripting language is necessary to use this Geant4-DNA simulation. For the future, physics, chemistry, damage repair kinetics, and survival function models will be refined and incorporated in the "molecularDNA" example. New DNA and cell geometries are being prepared to be included in the near future and give access to investigate systems with higher complexity. New models are also in development to implement interactions in other mediums than water. In addition, assemblies of cells will be investigated to access systems that imitate multicellular organisms or even human tissues in a well-defined radiation environment.

## Aknowledgements

We wish to thank Dr Mathieu Karamitros for his prototype contributions to the molecularDNA example. The authors also thank the European Space Agency for its support to Geant4-DNA through the "BioRad3" project (contract 4000132935/21/NL/CRS, 2021-2023).